\documentstyle[aps,pre,epsf]{revtex}
\begin{document}

\twocolumn[\hsize\textwidth\columnwidth\hsize\csname
@twocolumnfalse\endcsname

\title{Sequential fragmentation: The origin of columnar quasi-hexagonal
patterns}
\author{E. A. Jagla$^1$  and A. G. Rojo$^2$}
\address{$^1$Centro At\'omico Bariloche and Instituto Balseiro, Comisi\'on
Nacional de Energ\'{\i}a At\'omica\\
(8400) Bariloche, Argentina\\
$^2$ Department of Physics,
The University of Michigan\\ Ann Arbor, Michigan,
48109-1120, USA}
\maketitle
\begin{abstract}
We present a model that explains the origin and
predicts the statistical properties of columnar
quasi-hexagonal crack patterns, as observed in
the columnar jointing of basaltic lava flows. Irregular
fractures appear at the surface of the material, induced by
temperature gradients during cooling. At later times
fractures penetrate into the material, and tend to form polygonal
patterns. We show that this ordering can be described as a
tendency to minimize an energy functional. Atomistic simulations
confirm this interpretation.
Numerical simulations based on a phenomenological implementation of this
principle generate patterns that have remarkably good statistical
agreement with real ones.
%We find that the
%typical lateral size of the columns is proportional to the power
%$-2/3$ of the temperature gradient at which fractures form.
\end{abstract}

%\newpage

\vskip2pc] \narrowtext

\section{Introduction}

Columnar jointing in some kinds of volcanic
rocks--especially basaltic lava flows--is one spectacular example of
geometrical order in nature, where cracks split the rock in a set of
parallel columns \cite{holmes,walker}.
Perpendicularly to the columns
the fractures show a distinctive pattern
of mostly pentagonal and hexagonal polygons whose sizes vary from a few
centimeters to about two meters (see Fig. \ref{f1}). It has been
realized for more than a century that columns result from the
contraction of the cooling lava after solidification. There is
consensus by now that fractures start at the surfaces of the igneous
body, and propagate to the interior as the rock cools down. Fracture
patterns at the surface are rather disordered, but the fractures
progressively order as they penetrate the sample, reaching an
almost stable polygonal configuration after some depth. The
fundamental reason of this ordering process is unknown at present.
Further evidence for these facts come from the reproduction
of columnar cracking in samples of dessicating starches\cite{muller}. In
this case the columns have diameters in the range of the millimeters,
and the ordering process is apparent.

Ryan and Sammis \cite{ryan} collected evidence suggesting that the vertical
advance of the fracturing front is not continuous,
but a discrete process that we call {\em sequential fragmentation}:
at each step, and when some maximum tensile stress is exceeded,
a layer of material is fractured. The existence of this
punctuated advance of the fracturing front can be seen in
the lateral faces of the columns, which usually have typical
marks called striae, or chisel marks\cite{ryan}.
Further work\cite{degraff,aydin} has
clarified the form in which fractures propagate
within each horizontal layer. A fracture appears
at some point and propagates under the combination
of mode I and mode III fracturing, guided
by the upper part of the rock, which was fractured
previously. This is a justification
for the prismatic form of the columns and the
striae on their faces. However, it assumes that
the polygonal pattern is already formed in the upper parts of the
rock. Aydin and DeGraff\cite {aydin} tried to give a justification
for the evolution of superficial fractures, which
usually meet in the form of a `T', towards
the `Y' triple junctions typical of well developed columns.
But, as they point out, the prediction of the overall
polygonal pattern of fractures would require a three dimensional mechanical
analysis of the interaction among many neighbor
triple junctions. This is a extremely difficult task
if pursuit from a microscopic point of view.
They conclude that probably energetic
arguments (involving fracture energy and elastic energy)
may dictate the way in which the final
polygonal pattern is formed.

Energetic arguments have been invoked since quite a long time
to justify the polygonal structure of columnar basalts\cite{mallet}, and
it is known that the perfect hexagonal pattern
relieves the maximum amount of elastic energy
for a given total length of fractures\cite{foams}.
We argue in the next section
that for the problem of sequential fragmentation
(in which new parts of the
rock are sequentially fractured under the influence of both the previously
fractured parts and the still intact rock underneath)
a minimum principle can be invoked to describe
the evolution in depth of the fracture pattern. In the next
section we actually show
in a numerical simulation (which assumes only the sequential nature
of the fragmentation process and no other ad-hoc assumption)
that the fracture pattern
starts being disordered at the surface, and progressively
orders as it penetrates the sample, approaching the ideal pattern
we expect on an energetic basis.
The minimum principle is used in section III in phenomenological
simulations,
to evolve superficially disordered patterns into stable
polygonal configurations. The statistical
properties of the final patterns agree with the
available experimental data in basalts and also in starches.
In section IV we summarize our results, and point out some
open problems in columnar jointing, mainly associated to
realistic conditions of cooling.

\section{Energetic description of the ordering process}

We will concentrate on the problem of a semi-infinite solid body
(`the rock') cooling down through a (horizontal) free surface
\cite{equiv}. Since this is a situation of inhomogeneous cooling,
there will be thermal gradients within the rock. Thermal gradient
will point vertically at every point, and then
temperature will be constant in any
horizontal plane. Under the
stresses generated by the thermal gradient, the rock will fracture.

There are two qualitatively different stages
in the fracturing process. One is the appearance of
fractures at the surface of the rock. Here, the first fracture appears
when some maximum stress
is exceeded at some point of the sample. Then it propagates
horizontally under the influence of the inhomogeneities of
the rock. When new fractures nucleate at the surface, they propagate
until they meet older ones, usually at right angles, giving rise to
typical surface fragmentation patterns that have been extensively
studied, both experimentally\cite{exper,shorlin} and theoretically\cite{sokolov,teor}.
For our purposes, we only mention that this stage is governed to a large
extent by
the random disorder present in the system, since fractures
nucleate at points where the body can resist the lowest
strain. The pattern at the surface is usually quite disordered.

In this paper we study the second stage of the fracturing process
of the rock, namely the way in which the superficial, disordered
pattern of fractures penetrates the body and orders. We will
assume that the temperature distribution within the rock is a
given function of coordinates and time, independent of the actual
arrangement of fractures, and homogeneous at each horizontal plane.
The last fact, however, is not enough to assure that the fracture
front (i.e, the vertical coordinate up to which fractures have
penetrated, as a function  of the horizontal coordinate) will
be horizontal. In fact, in a standard situation of fracture
mechanics\cite{libro}, it would occur that as soon as a fracture penetrates
slightly more than the rest, stresses accumulate onto that fracture,
the result being that typically a single fracture advances.
For our case however, and under realistic cooling conditions,
the temperature gradient
decreases ahead of the fractures, so if a fracture advances,
it rapidly reaches regions where the lower temperature gradient
precludes the further advance of that fracture. This is the
reason for the sequential advance of the fracture front as
schematically illustrated in Fig. \ref{f2}.
At each ``time
step", a horizontal slab of material right below the fracture
front is fractured under the influence of the already fractured
material above, and the still unfractured material below. We will
always assume that the conditions for sequential fragmentation apply.

>From now on we will treat the rock as a
collection of particles, elastically joined
to their nearest neighbors, in the presence of
a constant temperature
gradient in the vertical direction.
Changes in temperature are interpreted as changes in the
equilibrium distance between particles. Fractures will be modeled by
the saturation of the elastic energy between neighbor particles as
they are taken apart a distance larger than some pre-fixed value
$d_{cr}(T)$ (see Fig. \ref{f3}). Note that, in this way, for any
given temperature distribution, and for any arrangement of particles,
we can define a total energy $E$ for the system without ambiguity.

It is useful to divide the total energy $E$ in two parts, $E=E_1+E_2$.
The $E_1$ term, that we call elastic energy, comes from those particles
being at relative distance lower than $d_{cr}$. This is an elastic
energy since is quadratic in the relative distance between particles.
The second part $E_2$ is the contribution from particles at relative
distance larger than $d_{cr}$, and then it can be associated with
broken links (since in this case force vanishes), and identified with
the fracture energy. Actually, whenever we talk of the existence of
a fracture at a given position of the sample, we mean that neighbor
particles are separated by a distance larger than $d_{cr}$ across
that `fracture'.

Having defined the degrees of freedom  of the system
and the total energy, we can think of the system as a point
$\cal P$ in the configuration space of all particle coordinates.
The sequential evolution we have described (Fig. \ref{f2}),
corresponds to the
sequential mechanical relaxation of all particles within the slab
between $z_i$ and $z_{i+1}$, with $dz=z_{i+1}-z_{i}$ being the
thickness of the slab being fractured at step $i$. This sequential
process corresponds to the movement of $\cal P$ in the energy landscape.
Assuming that the mechanical relaxation occurs by some kind of `viscous'
dynamics, the present description becomes complete and
deterministic, and then we can
solve in principle all the (non-linear) mechanical equations for the
problem, and obtain in all detail the way in which fractures penetrate
the sample. The qualitative features of this advance, however, can be
inferred from general arguments. In fact, with the fracture front at a
given $z$ position, we can calculate the stress field
ahead of the fractures, and determine the
directions along which this stress is maximized.
These are the directions that fractures have the tendency
to follow as they advance. The system releases the maximum
amount of energy when fractures advance along these directions,
compared to any other.
In other words, at each step the configuration point $\cal P$
moves following a steepest descendent
path in the potential energy landscape. Note that due to the
particular conditions of sequential advance, this movement is
`quasistatic', in the sense that it does not involve the runaway
of fractures ahead the fracture front.

The kind of argument we are using is equivalent to those used in surface
fragmentation to justify the fact that new fractures meet older ones at right
angles\cite{walker}. This is a consequence of the tendency of fractures
to advance perpendicularly
to the direction of maximum stress, and is equivalent to say that the
configuration point $\cal P$ moves down in energy following the steepest
descendent path. We are just saying that for sequential fragmentation the
advance of {\em all} fractures is governed by this kind of principle.
Then, our minimum principle, central to all this work, states that under
sequential fragmentation conditions the advance of the fracture front occurs
with a tendency to reduce as much as possible the total energy of the
system.
Note that during this ordering process, the existence of small inhomogeneities
in the material plays no significant role, as energy will be mostly dependent
only on the geometrical configuration of fractures.

Our principle then justifies qualitatively the observed tendency to produce
polygonal arrangements. It is important to note,
that the system finds the  most convenient pattern by
modifying the one at the surface (which usually is quite disordered) through
small steps as fractures penetrate the sample.
In the next two sub-sections we present results that confirm the validity
of our interpretation.

\subsection{A stress calculation}

First of all we want to show that standard stresses
calculations are consistent with the minimum principle.
We have calculated the stress field surrounding a system of unevenly
spaced fractures in two dimensions.
More specifically, we want to calculate the stress field for a set of
fractures as depicted in the inset of Fig. \ref{f4}, namely, there are pairs of
fractures separated by a distance $d_1$, and the pairs themselves are separated
by some other distance $d_2$.

We start with lattice points
joined by springs to form a triangular lattice, then modeling
an homogeneous and isotropic material with Poisson ratio \cite{ll} $\sigma=%
2/(4+\sqrt 3)\sim 0.35$.
%Lam\'e coefficients $\lambda$
%and $\mu$ satisfying $\lambda/\mu=4/\sqrt 3$.
We simulate a piece of size
$l_x\times l_z$ in the $x$ and $z$ directions respectively, taking
periodic boundary conditions in the $x$ direction, and open
boundary conditions in the $z$ direction.
The springs have a rest length $d_0$ that depends on its
vertical coordinate $z$ according to

\begin{equation}
d_0(z)=d_{00}(1-\beta\frac {z}{l_z})
\label{d0}
\end{equation}
In this way we model a constant temperature gradient in the $z$ direction (in
the simulations we will use $\beta=0.01$).
The periodic boundary conditions in the $x$ direction are taken in such
a way that the particles at $z=0$ are nominally at zero strain,
whereas all planes on top of that are strained with respect
to the preferred distance $d_0$.
The two fractures are introduced in the system by eliminating all springs
that go across the fractures.

We have solved numerically the problem, by relaxing (with a viscous dynamics)
the coordinates of the particles in order to obtain
the equilibrium configuration. Then the stress tensor\cite{ll} was calculated and
diagonalized at each position.
In Fig. \ref{f4}
we show the results. At each point, the tangent to the line shown in that
figure is the direction perpendicular to the eigenvector corresponding
to the maximum eigenvalue of the stress tensor, and then it is the direction
that fractures will tend to follow as they advance.
Starting at the
tips of the fractures, we see that these directions go away from each other, as
indicated by the arrows.
This indicates that, if sequential fragmentation occurs, the close fractures
will advance with a tendency to separate from each other, and eventually
to produce a set of evenly spaced fractures.
In fact, only when the evenly spaced configuration is reached,
the maximum stress direction will coincide
with the vertical direction, and from here the pattern is not modified.
This standard calculation coincides qualitatively with
what expected from the minimum principle, since
a set of evenly spaced fractures is the configuration that releases the
maximum amount of energy
(this is the equivalent of the honeycomb lattice in three dimensions\cite{dosd}).
Then we see that the conclusions from our minimum principle do not contradict
those obtained from more standard analysis. The advantage, however, is that the
minimum principle is much easier to implement in cases where a calculation
of stresses is not feasible.

\subsection{Atomistic simulations of ordering}

The second result presented to validate the minimum principle is an
atomistic numerical simulation in three dimensional systems.
We implement sequential fragmentation
in the following way. We use a generalization of
the procedure extensively used to study
surface fragmentation\cite{sokolov}.
In that case a layer of material shrinks while it
is attached to a fixed underlying layer.
We take an hexagonal plane of particles, with particles
attached to their neighbors by generalized springs (with an
energy-displacement relation as that of Fig. \ref{f3}) of spring constant
$K$ and initial
natural length  $d_0$. Their positions are the dynamical variables.
They are attached to an underlying hexagonal plane of particles (which are
kept fixed to their original positions during the simulations) by vertical
springs of constant $k$. The vertical springs do not break.
Simulation proceeds by reducing the equilibrium distance of the horizontal springs
of the layer being simulated.
The first fracture appears when the equilibrium distance between two particles
becomes grater than the corresponding critical distance $d_{cr}$ of the
spring that joins them.

For our simulations of sequential fragmentation, the only difference is that
we consider also the simulated plane to be joined to an upper plane of fixed
particles by spring of constant $k$, and that we simulate the fracturing
of the system as a sequence of independent two dimensional fragmentation
processes.
In the simulation of the successive layers the position of particles in the
upper plane are taken equal to
the final positions of the simulation of the previous layer.
In the simulation of the first layer, we do not have an upper plane. However,
to avoid introducing disorder into the system (and in order to break the homogeneity
that would occur for an absolutely perfect system) we take an upper plane
consisting of particles located at the hexagonal lattice plus some random displacement,
independent for each particle. We took this displacement to be 0.5 of the
lattice parameter. We want to mention that other simulations in which disorder was
included, and the first layer was simulated without any upper plane produced
qualitatively the same results. The equilibrium distance between particles $d_0$
within the layer
being simulated is quasistatically reduced from some initial value $d_{00}$ to
$pd_{00}$. We use $p=0.89$. In the energy of the horizontal springs
(Fig. \ref{f3}) we use $d_{cr}=d_0+0.1d_{00}$. We also take $K/k=100$.

In Fig. \ref{breaks} we see the final pattern of fractures for progressively
deeper layers $n$, for a system of $1600$ particles.
As we see, the pattern of fractures that appears is highly disordered for
the first few planes, with many fractures ending in the middle of the sample.
When we go inside the material, there is a clear tendency to order, forming
a polygonal pattern reminiscent of the experimental observations in basalts.
Although in Fig. \ref{breaks} some influence of the hexagonal structure
chosen for the underlying lattice is observable, we have verified that the
same qualitative process of ordering is found also for other underlying
geometries, namely square.
We have also looked at the final energy the pattern gets after fracturing,
and this quantity is plotted in Fig. \ref{eden} as a function of $n$.
As we see, this quantity has a tendency
to be minimized as successive layers are fragmented, which is the right tendency
predicted by our arguments. Moreover, in Fig. \ref{eden} we also plot the
energy expected for a perfect
hexagonal pattern, with the size of
the hexagons chosen precisely in order to minimize the energy. We see
that the solution that was found by the system was not the perfect one,
but very close in energy to that. This is a further confirmation that the
tendency to minimize the final energy is in fact the driving force for the
formation of the polygonal pattern.

\section{Phenomenological calculation}

Having identified the reason why a superficially disordered pattern shows
a tendency to order as it penetrates the material does not exhaust the
interesting features of the problem. Here we will address the observation that
patterns are usually seen to be polygonal, but not perfectly hexagonal, as it
would be preferred by purely energetic reasons. We will show that this is a
consequence of the minimization process, since the system is usually
not able to reach
the absolute minimum of the energy potential, but gets trapped in
a relative minimum.
Since the problem becomes computationally too costly to be tackled by the
methods of the precedent section, we look for a phenomenological approach.
We will need to calculate in some
approximate manner the energy of the system as fractures advance, in order to
search for fracture patterns that tend to minimize the energy.

A realistic calculation is rather complicated and it will be presented
elsewhere. Here we will restrict to an heuristic analysis that however
is able to show many of the known physical properties of fracture patterns.

Let us suppose that fractures divide the system in sectors of well defined
areas $A_i$. We are interested in the elastic energy $E_1$ of the system after
a vertical advance $dz$ of the fractures. To lowest order this energy must
be a function of the $A_i$, of the elastic constants, and of the
precise thermal state of the material.
We will use the following expression
\begin{equation}
E_1=E_0 +\gamma \sum_i A_i^\nu dz,
\label{ener2}
\end{equation}
where $\gamma>0$ and $\nu>1$ are constants, and we have collected within $E_0$ all
possible terms that do not depend on $A_i$. Three main facts
have been used in constructing
expression (\ref{ener2}).
First, the energy is an independent sum over different columns of terms that depend
on $A_i$. This is the lowest order contribution we expect, in which we disregard
contributions proportional to the particular form of the columns,
and interaction terms
between neighbor columns. Second, the final energy $E_1$ increases if $A_i$
increases (i.e., $\gamma>0$).
This is the right tendency, since
the final elastic energy becomes lower if new fractures are introduced in
the system, and this implies a reduction of the typical $A_i$.
Third, the exponent $\nu$ must be greater than one. This condition
implies the tendency
of the system to make the distribution of $A_i$ as uniform as possible in
order to reduce $E_1$. With illustrative purposes,
in the rest of this paper we will use $\nu=2$. We have repeated the simulations with $\nu$ in the 1.5 to
2.5 range with no significant change.
The precise properties of the material an the thermal state of the system
are contained in the value of $\gamma$.

Expression (\ref{ener2})
for the final elastic energy has to be added with
the change in the fracture energy $E_2$ during the vertical advance.
This is simply given
in terms of the energy needed to create the new fractures as

\begin{equation}
\delta E_2=\eta L d z,
\label{ener}
\end{equation}
where $\eta$ is the fracture energy per unit area, and $L$ is the total length
of fractures perpendicular to the propagation direction.
Collecting the elastic (\ref{ener2}) and fracture (\ref{ener})
energy terms, we can rephrase the minimum principle
in the following form. Upon fracture advance, the energy functional
\begin{equation}
{\cal E}=\gamma \sum_i A_i^2 +\eta L
\label{energ4}
\end{equation}
tends to be minimized.
The absolute minimum of (\ref{energ4}) is attained by a
perfect pattern of hexagons of side

\begin{equation}
l_{\min }=\left( 2\eta /9\gamma\right) ^{1/3}
\label{lmin}
\end{equation}
%This formula expresses the typical size of the polygons in terms of elastic
%coefficients of the rock and the thermal gradient under which fractures
%form. To our knowledge this is the first closed
%prediction of the stable size of the polygons
%in terms of parameters of the material, and the typical temperature gradient
%under which they form.

Now, we will use  functional (\ref{energ4})
to evolve irregular patterns (representing
superficial fractures) up to point in which they stabilize, and then compare
their statistical properties with real ones.
Since we are not able to manage a completely general case,
we chose a simple possibility
that turns out to produce quite interesting results.
We generate the
pattern at the surface by a process of nucleation of linear
fractures: from randomly chosen points within the plane we propagated two
opposite, straight fractures. The process was repeated many times, with new
fractures stopping as soon as they reached an older fracture. In Fig. \ref
{fig2}(a) we show a typical pattern generated by this process\cite
{tjunctions}. We simulate the modification of the pattern with an algorithm
that makes small changes to the positions of the nodes at which fractures
join. Each step in the modification of the pattern corresponds to the
fracture pattern developing into the rock. The new position for a node was
accepted if the new value of the energy, as given by Eq. (\ref{energ4})
was lower than the previous value. In
addition, at each step of the simulation the configuration was checked for
the existence of very close nodes that can allow a change in the topology of
the pattern according to the sketch of Fig. \ref{fig2}(d). Again, the
changes were accepted only if they reduce the value of $\cal E$. These processes
are important since they change the number of sides of the polygons, and
allow for a progress towards more stable patterns.

An intermediate pattern in the evolution process is shown in Fig. \ref{fig2}%
(b), and the final one (after which all proposed changes of the positions of
the nodes increase the energy) is shown in Fig. \ref{fig2}(c).
Since `time'
on our simulations corresponds to `depth' in the rock, the ordering of our
patterns represents the progressive order of the real lava fractures deeper
into the rock \cite{ojo}. The final pattern of Fig. \ref{fig2}(c) is not
perfectly hexagonal,
and thus it is only a relative minimum of (\ref{energ4}). There is one
single effective parameter in the simulation, that can be taken to be
the side of the perfect hexagonal pattern of minimum energy $l_{\rm min}$.
For our simulations this value, as given by (\ref{lmin}),
is indicated in Fig. \ref{fig2}.
The qualitative similarity of the final pattern with that of the Giant's,
Causeway shown in Fig. \ref{f1}, is apparent.
This polygonal pattern is now exposed at the surface of the
rock, but there is evidence that this is not the original surface. In Fig.
\ref{estad} we show two quantities that are a measure of the statistical
similarity between our patterns and the real ones. In Fig. \ref{estad}(a) we
see the results for the frequency of appearance of polygons with a given
number of sides (in this case we also include the results on cornstarch by
M\"{u}ller\cite{muller}), and in Fig. \ref{estad}(b) the corresponding
values for the mean area of polygons with a given number of sides, both in
our simulations and in the real patterns. The configurations generated by
our model are remarkably realistic. We see that, both in real cases and in
our simulations, the fractures never reach a perfect hexagonal pattern.
Instead, a reproducible distribution of polygons, most of them with 5, 6,
and 7 sides is obtained, with a minor contribution of polygons with 4 and 8 sides.
Also, polygons with higher number of sides have larger area as Fig. \ref
{estad}(b) shows.
%We have also included in Fig. \ref{estad} the results from
%Ref. \cite{budke} which is (to our knowledge) the only previous attempt to make
%a quantitative prediction of statistical properties of the patterns. We see that
%their distribution is unrealistically peaked at $n=6$, and the size
%distribution does not show the right tendency. Our results are clearly more
%realistic.

\section{Summary and perspectives}

In this paper we have given a first approach to a consistent model for the
existence of columnar polygonal patterns in lava flows and some dessicating
materials. We have shown in numerical simulations on a discrete model
that fractures
appear as irregular cracks at the free surface of the material and become
ordered as they penetrate into the interior. We have argued that this effect
is a consequence of a tendency to minimize an energy functional. The process
of minimization follows a rough landscape, and is always towards a local minimun.
This process is therefore monotonically decreasing with the thermal fluctuations playing no role.
Relying on
this principle, we showed that the statistical properties of experimental
polygonal patterns can be reproduced.

There are still some problems that deserve further consideration and that
we plan to discuss in a forthcoming publication. They have to do
mainly with the
realistic conditions of cooling. As discussed in section II, it is precisely
the decreasing of the temperature gradient ahead of the fractures that
makes possible the sequential advance of the fracture front, in
a coordinated way all across the sample. The detailed study of this problem
provides predictions for the width of chisel marks on the columns.

We also have to determine in a realistic situation the value of the constant
$\gamma$ and $\nu$ in expression {\ref{ener2}. This will allow to calculate in
particular the typical values for the polygons in basalts and starches.
Under realistic cooling conditions we also have to face the problem that
temperature changes with time, and the effect of this on the advance of
the fracture front has to be discussed.
%We will show elsewhere than (\ref{lmin}) has to be complemented with another
%expression that gives the necessary conditions for the advance of the pattern.
%This expression can be obtained from the application of a sort of
%Griffits criterium to our particular problem.
%The analysis allows to make a discussion of what are the parameters that
%fix the size of the polygons under realistic conditions of cooling. We find that
%at the end, the size of the poligons is mainly controlled by the size of the
%irregular patches already present at the surface.

We thank Roy Clarke, Eric Clement, Alan Cutler, Eric Essene, Len Sander and
Youxue Zhang for very useful suggestions. A.G.R. acknowledges partial
support from the National Science Foundation. E.A.J. acknowledges the hospitality
of ICTP, Trieste, Italy, where part of this worked was done, and financial support
from CONICET (Argentina).

\begin{figure}
\narrowtext
\epsfxsize=3.3truein
\vbox{\hskip 0.05truein
\epsffile{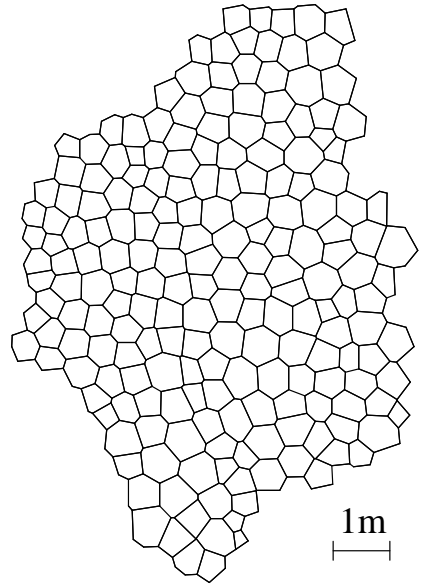}}
\caption{Polygonal pattern seen perpendicularly to some of the
columns of  the Giant's Causeway, a Tertiary lava flow in Antrim, Northern
Ireland, from Ref. \protect\cite {budke} (originally from a map by
O'Reilly\protect\cite{oreili})).}
\label{f1}
\end{figure}

\begin{figure}
\narrowtext
\epsfxsize=3.3truein
\vbox{\hskip 0.05truein
\epsffile{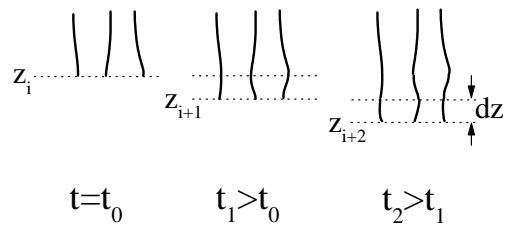}}
\caption{Schematic representation of sequential fragmentation
in a two-dimensional geometry . Continuous lines represent
the advancing fractures. At each time step a layer of material of
thickness $dz$ fractures. New fractures appear below those already
present, but slight modifications in their positions are possible,
and in fact crucial to the ordering process.}
\label{f2}
\end{figure}

\begin{figure}
\narrowtext
\epsfxsize=3.3truein
\vbox{\hskip 0.05truein
\epsffile{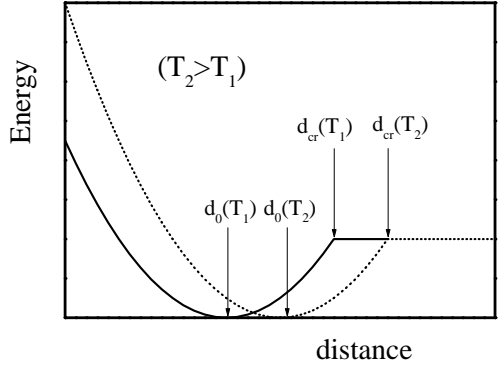}}
\caption{Schematic energy vs. distance curve for neighbor particles in the
discrete model of the rock. There is an equilibrium distance between
particles that depends on temperature $d_0(T)$. Deviations
from this distance costs an elastic energy which is quadratic
in the displacement. If the distance becomes greater than some
critical distance $d_{cr}(T)$ then energy saturates. In this way
we model fractures, since in this range there is no force between
particles. }
\label{f3}
\end{figure}

\begin{figure}
\narrowtext
\epsfxsize=3.3truein
\vbox{\hskip 0.05truein
\epsffile{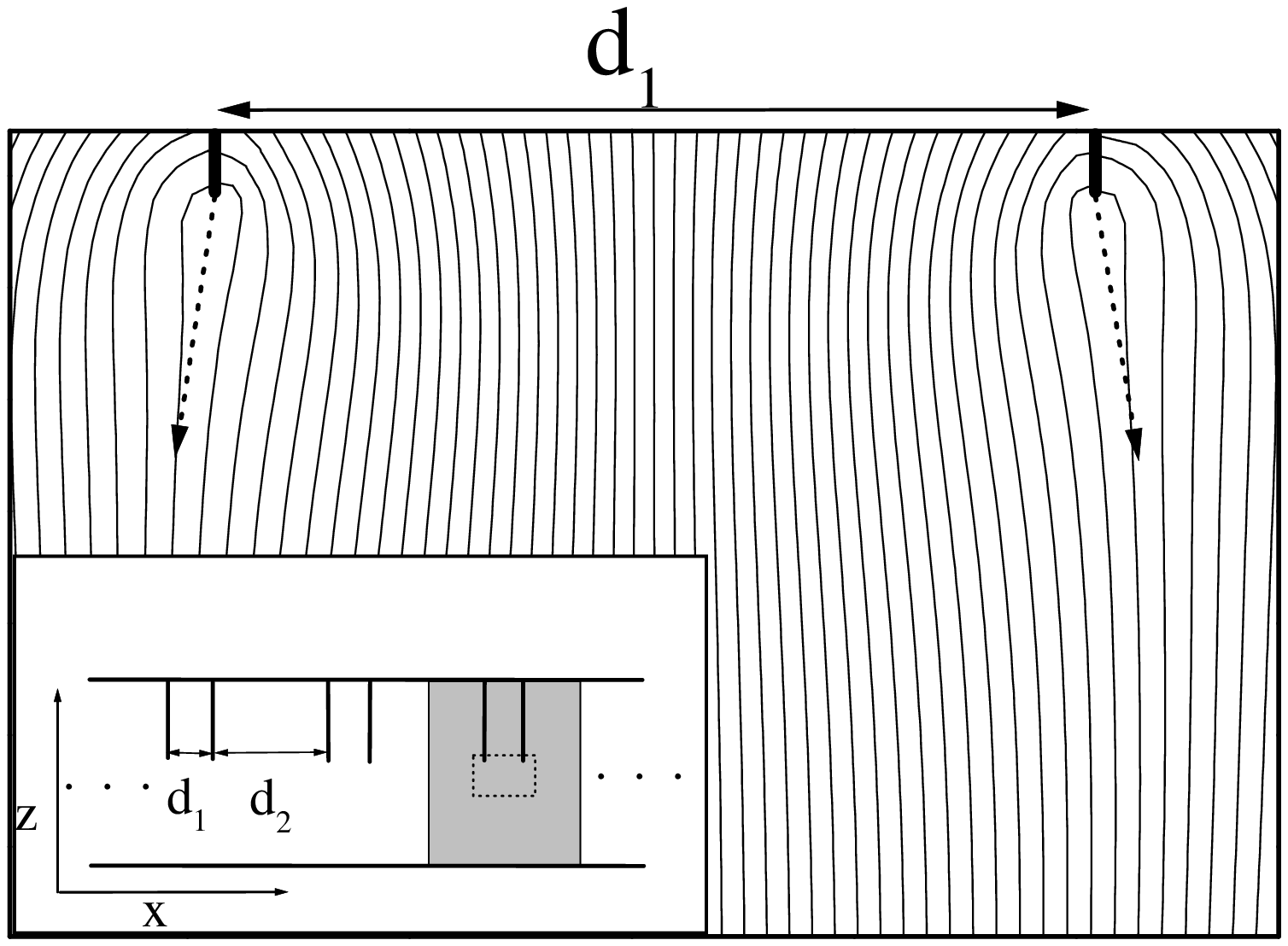}}
\caption{Stresses ahead of an unevenly set of fractures as depicted in the inset.
In the main figure we plot the stress field in the neighborhood of a couple
of close fractures (see text for details).
Directions of maximum stress at the tips of the fractures are
indicated by the arrows.
The simulated box is marked in gray in the inset. Periodic
boundary conditions are used along $x$, and free boundary conditions along $z$. The
dotted box in the inset is the region plotted in the main figure. We have used
$d_2/d_1=4$.}
\label{f4}
\end{figure}

\begin{figure}
\narrowtext
\epsfxsize=3.3truein
\vbox{\hskip 0.05truein
\epsffile{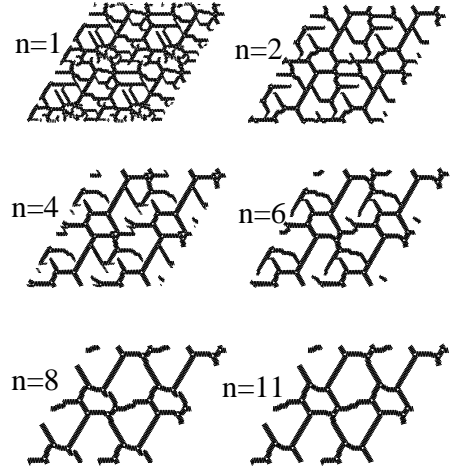}}
\caption{The final patterns of fractures for progressively deeper layers $n$
in a sequential fragmentation process for layers with 1600 particles. For clarity
reasons, in the plots  the system has been duplicated both in horizontal and
vertical direction (periodic boundary conditions are used). In the plots,
each thick fracture is formed by small lines mostly perpendicular to the
fracture that join the ends of springs that have failed after a contraction
up to $0.89$ of the original distance between lattice points.}
\label{breaks}
\end{figure}

\begin{figure}
\narrowtext
\epsfxsize=3.3truein
\vbox{\hskip 0.05truein
\epsffile{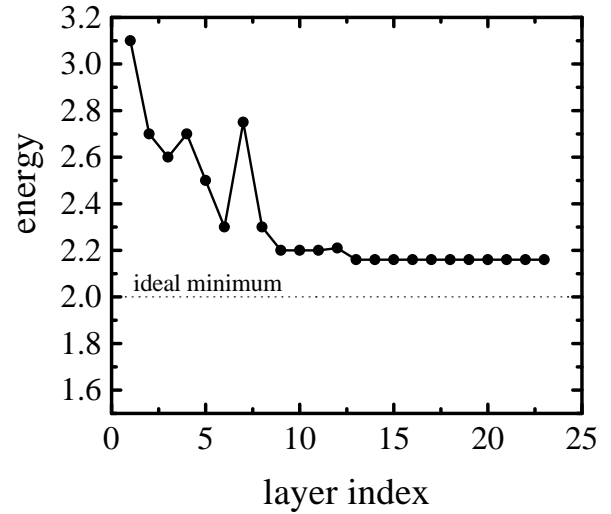}}
\caption{Energy per particle of the pattern obtained in the simulations shown in the
preceding figure, as a function of the layer index. The ideal minimum of
a size-optimized honeycomb lattice is also shown. Energy is given in units
of $10^{-3}Kd_{00}^2$.}
\label{eden}
\end{figure}

%\begin{figure}
%\narrowtext
%\epsfxsize=3.3truein
%\vbox{\hskip 0.05truein
%\epsffile{quilombo.eps}}
%\caption{(a) Elastic energy of a column of a given area in the presence
%of a thermal gradient, before and after a vertical advance of the fracture front
%by $dz$. The relieved energy starting from a fixed point of the `before' curve
%becomes greater as the final area $A'_i$ becomes lower, as indicated
%($\delta E'_1>\delta E_1$). The sketch of the configurations before and after the
%advance are shown in (b).}
%\label{quilombo}
%\end{figure}

\begin{figure}
\narrowtext
\epsfxsize=3.3truein
\vbox{\hskip 0.05truein
\epsffile{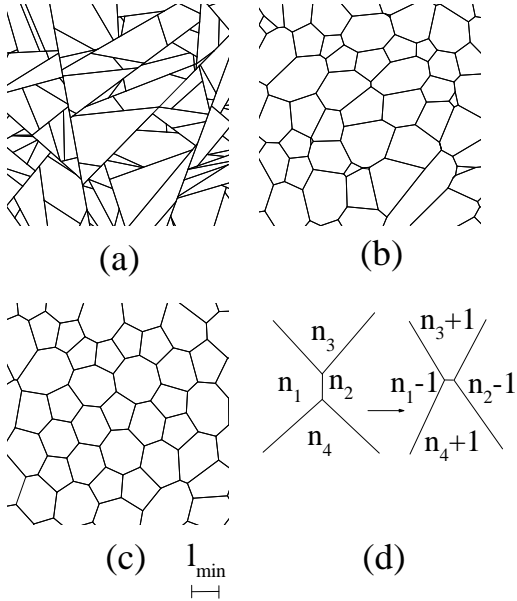}}
\caption{Numerically evolved patterns of fractures. We see the original
pattern (a), the final (stable) one (c), and one intermediate configuration
(b). To avoid spurious edge effects, only the central region of a simulation
performed on a larger sample is shown. The numerical algorithm is described
in the text. Below the final pattern (c), the side of hexagons $l_{\rm min}$
in the expected ideal honeycomb lattice is indicated.
In (d) we see the kind of processes that allow for a change in
the number of sides of adjacent polygons.}
\label{fig2}
\end{figure}

\begin{figure}
\narrowtext
\epsfxsize=3.3truein
\vbox{\hskip 0.05truein
\epsffile{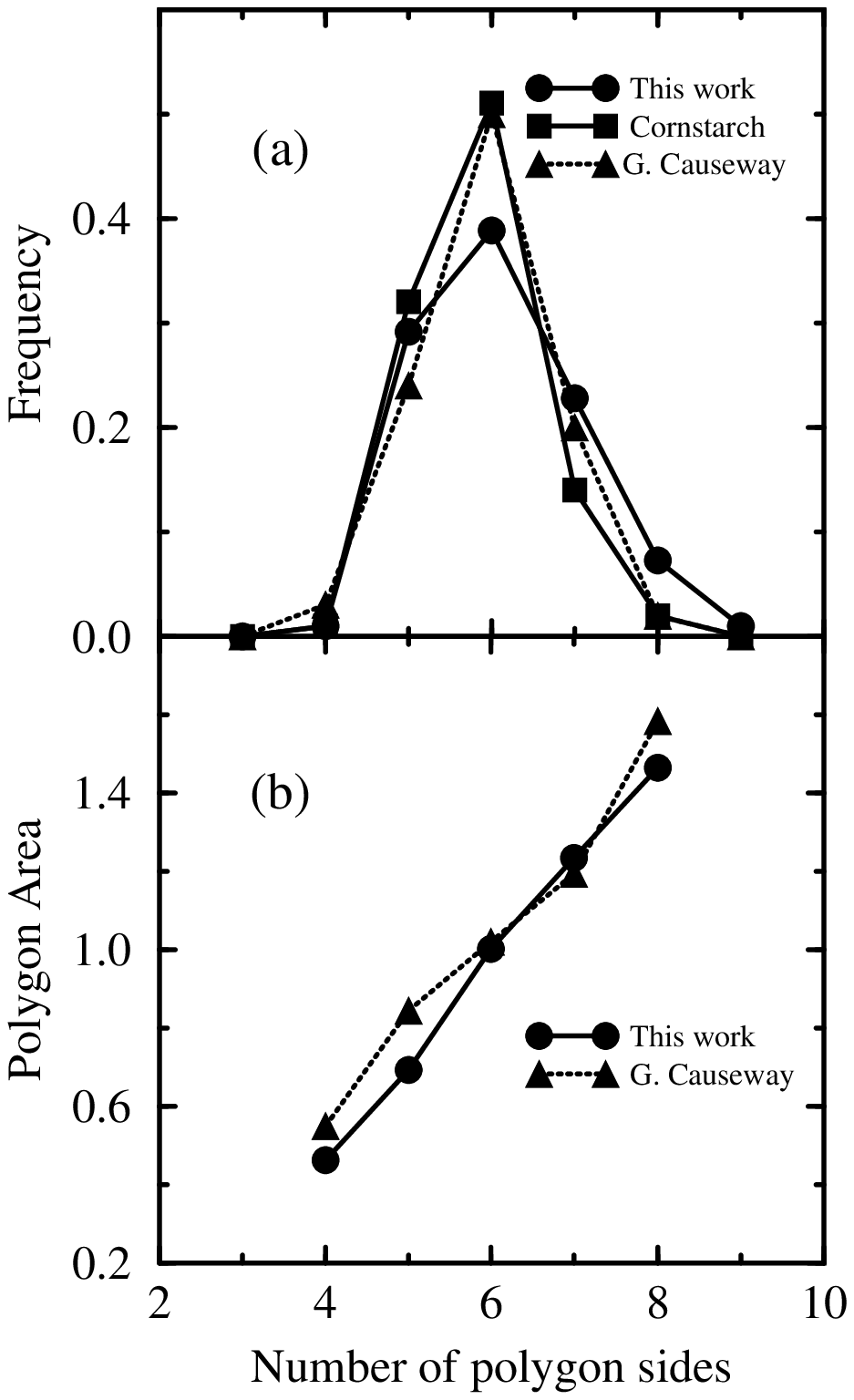}}
\caption{(a) Histogram for the relative frequency (normalized to one) of appearance of polygons with different number
of sides for the Giant's Causeway, columns in cornstarch\protect\cite{muller},
and from our simulations (an average over 10 final configurations as that
of Fig. \ref{fig2}(c) is shown). (b) Areas of the polygons normalized to the
average area for polygons with different number of sides (data on cornstarch
are not available).}
\label{estad}
\end{figure}


\begin{references}
\bibitem{equiv}We will use the language appropriate for the fracturing problem
under thermal gradients. The arguments, however, apply also to the
dessication problem in starches and other materials, by replacing `temperature' by
`humidity concentration' and similarly with other expressions.
\bibitem{dosd}In two dimensional dessication experiment, an evenly spaced set
of fractures is usually obtained (see {\protect\cite{allain}} and
{\protect\cite{shorlin}}).
Our minimum principle gives a natural justification to this fact.
\bibitem{tjunctions}  This procedure exaggerates the disorder in the first
layer. We have repeated our simulations starting with a pattern consisting
of 90 degrees junctions only, with identical results.
\bibitem{ojo}  In order that the final pattern be independent of the initial
density of fractures, this has to be high enough. This is because we do not
allow for the creation on new fractures in the algorithm, all of them have
to be there `from the beginning'.
\end{references}
\end{document}